\documentstyle[aps,preprint]{revtex}

\begin{document}

\hfill
{\vbox{
\hbox{CPP-95-13}
\hbox{DOE-ER-40757-070}
\hbox{UCD-95-24}
\hbox{September 1995}}}

\begin{center}
{\large \bf Heavy Quark Fragmentation Functions for \\
D-wave Quarkonium and Charmed Beauty Mesons}

Kingman Cheung\footnote{Internet address: {\tt cheung@utpapa.ph.utexas.edu}}

{\it Center for Particle Physics \\ University of Texas at Austin,
Austin, TX 78712 U.S.A.}

Tzu Chiang Yuan\footnote{Internet address: {\tt yuantc@ucdhep.ucdavis.edu}}

{\it Davis Institute for High Energy Physics \\
University of California at Davis, Davis, CA 95616 U.S.A.}
\end{center}

\begin{abstract}
At the large transverse momentum region, the production of
heavy-heavy bound-states such as charmonium, bottomonium, and $\bar
bc$ mesons in high energy $e^+e^-$ and hadronic collisions is
dominated by parton fragmentation. We calculate the heavy quark
fragmentation functions into the D-wave quarkonium and $\bar bc$
mesons to leading order in the strong coupling constant and in the
non-relativistic expansion.  In the $\bar b c$ meson case, one set of
its D-wave states is expected to lie below the open flavor
threshold. The total fragmentation probability for a $\bar b$
antiquark to split into the D-wave $\bar b c$ mesons is about $2
\times 10^{-5}$, which implies that only 2\% of the total
pseudo-scalar ground state $B_c$ comes from the cascades of these
orbitally excited states.
\end{abstract}

\thispagestyle{empty}

\section{Introduction}

The production of heavy quark-antiquark bound states like
charmonium, bottomonium, and the yet undiscovered charmed beauty
mesons at high energy $e^+e^-$ and hadronic machines can provide very
interesting tests of perturbative QCD. In the charmonium case,
our traditional wisdom has been taken the following scenario:
most of the $J/\psi$ comes predominately from either the radiative
decay of the P-wave $\chi_{cJ}$ states produced by the lowest order
gluon fusion mechanism or the weak decays of B mesons,
whereas the $\psi'$ should be produced almost entirely from
B meson decays.   Similarly, most of the $\Upsilon$ should be produced from
the radiative $\chi_{bJ}$ decay. Only until recently can
this naive thinking be confronted by the wealth of
high energy experimental data collected at the Tevatron.
During the 1992-1993 run, the CDF detector
recorded surprisingly large production rates of the prompt
$J/\psi$, $\psi'$, and $\chi_{cJ}$ which are orders of magnitude
above the theoretical lowest order predictions \cite{CDF}.
In 1993 \cite{gswave},
it was realized that fragmentation can play an important role in
quarkonium production at the large transverse momentum $(p_T)$ region
at the Tevatron. Parton fragmentation into quarkonium
is formally a higher order process but nevertheless it can be
enhanced at the sufficiently large transverse momentum region
compared with the usual gluon fusion process \cite{gswave}.
A simple explanation of the enhancement
is that the fragmenting partons are produced at high
energies but with small invariant masses which can lead to the enhancement
factor of powers of $(p_T/m_c)^2$ compared with the gluon fusion mechanism.

Most of the fragmentation functions for quarkonium that are relevant to
phenomenology have now been calculated. Gluon fragmentation
functions were calculated in Ref.\cite{gswave}
for the S-wave, Ref.\cite{gpwave1,gpwave2} for the P-wave,
and Ref.\cite{gdwave} for the $^1D_2$ state; heavy quark fragmentation
functions were calculated in Ref.\cite{qswave1,qswave2} for the S-wave, and
Ref.\cite{qpwave1,qpwave2,qpwave3} for the P-wave.
Phenomenological applications of these fragmentation functions have
been performed by a number of groups \cite{pheno1,pheno2}.
The current scenario of charmonium production at high transverse
momentum is as follows:
among all fragmentation contributions that are relevant to $J/\psi$
production, the dominant ones come from (i) gluon fragmentation into a
color-octet $^3S_1$ $c \bar c$ state followed by a double E1 transition
into $J/\psi$, and (ii) gluon fragmentation into $\chi_{cJ}$ followed by
the radiative decay $\chi_{cJ}\to J/\psi + \gamma$.
The former mechanism is the most dominant one for $\psi'$
production because of the absence of $\chi_{cJ}(2P)$ states below the
$D \overline D$ threshold. The current
experimental status with this new fragmentation insight has now been
reviewed by many people in various occasions.
We refer the readers to Refs.\cite{pheno1,pheno2,talks,braaten-talk}
for more details.

The charmed beauty $\bar b c$ mesons are expected to be
observed soon at the Tevatron and preliminary limits on the production rate
already exists \cite{cdf-bc}. Parton fragmentation has been applied
phenomenologically to the production of $\bar bc$ mesons as well at the
Tevatron \cite{cheung,induce}. The heavy quark fragmentation
functions for $\bar b^* \to (\bar bc)$ were calculated in Ref.~\cite{bcswave}
for the S-wave and in Refs.~\cite{qpwave1,qpwave2} for the P-wave.
The validity of the parton fragmentation has been questioned in some recent
exact ${\cal O}(\alpha_s^4)$ calculations \cite{bcdebate}. However,
these calculations obtained different results and led to conflicting
conclusions among themselves.
It was only until the recent appearance of the
results of Ref.\cite{leike} that this controversy
can be settled down and the validity of the fragmentation approximation
can be established in the production of $(\bar bc)$ mesons at the Tevatron.
One can, therefore, use the fragmentation approximation with confidence
in calculating the production cross sections  and transverse momentum spectra
of the $(\bar bc)$ mesons, including the leading logarithmic corrections,
induced gluon fragmentation contribution \cite{induce},
and contributions from all orbitally excited states.

The purpose of this paper is to extend the previous calculations of
heavy quark fragmentation functions to the D-wave case. The D-wave
orbitally excited states are of interests
phenomenologically. In the $(\bar b c)$ case, one set of the D-wave states
is predicted by potential models to lie below the $BD$ threshold. These
states, once produced, will cascade into the pseudo-scalar ground state
$B_c$ via pion and/or photon emissions and thereby contribute to the
inclusive production rate of $B_c$. The excited D-wave charmonium
resonances have also been suggested \cite{chowise}
to resolve the $\psi'$ surplus problem observed at CDF. In Section II,
we briefly review the factorization model of Bodwin, Braaten,
and Lepage \cite{bbl} that can consistently factor out
the long-distance physics and short-distance perturbative factors for
the inclusive production and decay rates of heavy quarkonium.
In Section III, we present the calculation
of the D-wave fragmentation functions for both the unequal and equal mass
cases within the spirit of the factorization model. We discuss our results in
Section IV and conclude in Section V.

\section{The Factorization Model}

A rigorous theory based on non-relativistic QCD (NRQCD) has been developed
recently by Bodwin, Braaten, and Lepage \cite{bbl} for the inclusive decay
and production of heavy quarkonium. Using this formalism, the fragmentation
function for a heavy quark $Q$ to split into a quarkonium state $X$
with longitudinal momentum fraction $z$ can be written as
\begin{equation}
D_{Q \to X}(z,\mu) = \sum_n d_n(z,\mu)  \langle {\cal O}_n^X \rangle \; ,
\label{factor}
\end{equation}
where ${\cal O}_n$ are local 4-fermion operators
defined in NRQCD.
The short-distance coefficients $d_n(z,\mu)$ are independent of the
quarkonium state $X$. For a fragmentation scale $\mu$ of order of the heavy
quark mass $m_Q$, the coefficients $d_n(z,\mu)$ can be calculated
using perturbation theory in strong coupling constant
$\alpha_s(2m_Q)$.  The relative size of the matrix elements
$\langle {\cal O}_n^X \rangle$ for a given state $X$
can be estimated by how they scale with $m_Q$ and with
the typical relative velocity $v$ of the heavy quarks inside the quarkonium.
Thus, the factorization formula Eq.(\ref{factor}) is a double expansion
in $\alpha_s$ and $v$.  To determine the relative importance of the various
terms in this formula, one should take into account both the scaling
in $v$ of the matrix elements $\langle {\cal O}_n^X \rangle$ and the
order of  $\alpha_s$ in the coefficients $d_n(z,\mu)$.
The leading term in $v$ of this formula corresponds
to the popular color-singlet model \cite{schuler}. However,  keeping only
the leading term in this double expansion can sometimes lead to
incomplete or even inconsistent results due to the
presence of infrared divergences \cite{bbl,bbl0}.

One important feature in the factorization model is that the
quarkonium state $X$ is no longer considered as solely a
$Q \overline Q$ pair but rather a superposition of Fock states. For example,
the spin-triplet $^3D_J$ quarkonium states, denoted by $\delta_J^Q$,
have the following Fock state expansion,
\begin{equation}
\vert \delta^Q_J \rangle  =
O(1) \vert  Q {\overline Q} (^3D_J,{\underline 1}) \rangle +
O(v) \vert  Q {\overline Q} (^3P_{J'},{\underline 8}) g \rangle
+  O(v^2) \vert  Q {\overline Q} (^3S_1,{\underline 8}\; {\rm or}\;
{\underline 1} ) gg  \rangle + \cdots
\label{fock}
\end{equation}
where the notations $\underline 1$ and $\underline 8$ refer, respectively,
to the color-singlet and color-octet states of the $Q \overline Q$ pair.
In the above Fock-state expansion there are also other ${\cal O}(v^2)$ states,
e.g., $\vert Q{\overline Q} (^3D_J, {\underline 8} \; {\rm or}\;
{\underline 1} ) gg \rangle$, but their production will be further suppressed
by powers of $v$.
Heavy quark fragmentation into
D-wave quarkonium $\delta^Q_J$ can be deduced from the leading
Feynman diagrams of
$Q^* \to (Q {\overline Q})Q$ as shown in Fig.1 and Fig.2.  In Fig.1,
the $Q \overline Q$ pair can be either in the color-singlet $^3D_J$ state,
in the color-octet $^3P_{J'}$, or in the color-octet or color-singlet
$^3S_1$ states,
whereas in Fig.2 it can only be in the color-octet $^3S_1$ state.
The short-distance factors deduced from these diagrams
are all of order $\alpha_s^2$. Furthermore, all these diagrams are also of the
same order in $v$.    According to the
NRQCD \cite{bbl}, the relevant local 4-fermion operators describing these
three Fock states are ${\cal O}_1(^3D_J)$, ${\cal O}_8(^3P_{J'})$, and
${\cal O}_8(^3S_1)$ which have relative scalings like $v^4$, $v^2$,
and $v^0$, respectively.
However, the $Q \overline Q$ pair in the color-octet P-wave
and S-wave states can evolve nonperturbatively into the physical
D-wave quarkonium state $\delta^Q_J$ by emitting one and two soft gluon(s),
respectively.  Emitting a soft gluon costs a factor of $v$ at the amplitude
level in NRQCD and hence a factor of $v^2$ in  the probability.
Thus, with a non-relativistic normalization convention in the state,
all three matrix elements
$\langle {\cal O}_1^{\delta^Q_J}(^3D_J) \rangle$,
$\langle {\cal O}_8^{\delta^Q_J}(^3P_{J'}) \rangle$, and
$\langle {\cal O}_8^{\delta^Q_J}(^3S_1) \rangle$ scale like $m_Q^3v^7$.
As a result, both Fig.1 and Fig.2 contribute to the fragmentation functions of
order $\alpha_s^2 v^7$ and should all be taken in account for a consistent
calculation.
With appropriate normalization, the color-singlet matrix elements
$\langle {\cal O}_1^{\delta^Q_J}(^3D_J) \rangle$
as well as
$\langle {\cal O}_1^{\delta^Q}(^1D_2) \rangle$ for the
spin-singlet D-wave state $\delta^Q$ can be related
to the non-relativistic radial D-wave wave-function
${\overline {R_D^{''}(0)}}$,
which is the spin average of the spin-singlet and spin-triplet
$R^{''}_D(0)$'s, by
\begin{equation}
\langle {\cal O}_1^{\delta^Q}(^1D_2) \rangle
\approx \frac{75N_c}{4 \pi} \vert {\overline {R_D^{''}(0)}} \vert^2 \; ,
\label{1d2_op}
\end{equation}
\begin{equation}
\langle {\cal O}_1^{\delta^Q_J}(^3D_J) \rangle
\approx  \frac{15(2J+1)N_c}{4 \pi} \vert {\overline {R_D^{''}(0)}} \vert^2 \; .
\label{3dj_op}
\end{equation}
Potential models can be used to determine the value of the
wave-function so that the NRQCD matrix elements for the color-singlet
contributions are fixed.
Unfortunately, potential models cannot determine
the color-octet matrix elements since dynamical gluons are involved.
However, from our experience of the P-wave quarkonium case \cite{qpwave1},
we do not expect the color-octet component to play a major role in the
heavy quark fragmentation. We note that this is in sharp
contrast with the gluon fragmentation in which a
gluon can fragment into the color-octet
$^3S_1$ state via the process $g^* \to Q \overline Q$. This process is
of order $\alpha_s$ and is at least one power of $\alpha_s$ lesser than the
leading color-singlet term in the gluon fragmentation.
We also note that the color-singlet contribution of heavy quark
fragmentation from Fig.1 is free of infrared divergences. On the other hand,
infrared divergences will show up in the gluon fragmentation function
into the spin-triplet D-wave quarkonium. It is necessary to include
the color-octet contributions in order to achieve a finite and sensible
perturbative answers for the gluon fragmentation functions into
spin-triplet D-wave states.  In what follows we will restrict ourselves to the
color-singlet contribution in the heavy quark fragmentation into D-wave
quarkonium.
If the color-octet matrix elements
can be determined in the future, it is straightforward to include
their contributions since
their corresponding short-distance coefficients can be extracted easily
{}from the previous S-wave and P-wave calculations by modifying the
color factors.

The factorization model for the quarkonium system can be extended to the
unequal mass case like the $(\bar b c)$ meson system. In what follows,
we will denote the spin-singlet and the spin-triplet D-wave
$(\bar b c)$ meson states by $\delta^{bc}$ and $\delta^{bc}_J$, respectively.
In this case, figure~2 is absent for the color-octet contribution.

\section{Heavy Quark Fragmentation Functions into D-wave
Heavy-Heavy Mesons}

The general covariant formalism for calculating the production and decay
rates of S-wave and P-wave heavy quarkonium in the non-relativistic expansion
was developed some times ago \cite{kuhn}.  It is straightforward to extend
it to the case of unequal mass and higher orbital excitation.
We shall present a covariant formalism for the production of D-wave meson
$(Q\bar q)$ by some unspecified short-distance processes.
Let $m_Q$ and $m_q$ be the masses of the two quarks with $m_Q > m_q$,
and introduce the
mass parameters $r=m_q/(m_Q + m_q)$ and $\bar r = m_Q/(m_Q + m_q) = 1 - r$.
In the leading  non-relativistic approximation, the mass $M$
of the meson is simply $m_Q+m_q$.
The amplitude for producing the bound-state $(Q\bar q)$ in a state with
momentum $P$, total angular momentum $J$,
total orbital angular momentum $L$, and total spin $S$
is given by
\begin{equation}
A(P) =\sum_{L_Z,S_Z} \int \frac{d^3 {\bf k}}{(2\pi)^3} \; \Psi_{L L_Z}(
{\bf k} ) \; \langle LL_Z;S S_Z | J J_Z \rangle \; {\cal M}(P,k)  \;,
\end{equation}
where
\begin{equation}
\label{covar}
{\cal M}(P,k) = {\cal O}_\Gamma \; \Gamma_{S S_Z} (P,k) \;.
\end{equation}
${\cal O}_\Gamma$ represents the short-distance interaction producing
the $Q$ and $\bar q$ in a specific relative orbital angular momentum,
and in general is a product of Dirac matrices.
The heavy quarks $Q$ and $\bar q$ have momentum $r P + k$ and
$\bar r P - k$, respectively,
where $2k$ is the small relative momentum of the quarks inside
the bound state.
$\Psi_{LL_Z} (\bf k)$ is the Bethe-Salpeter wave-function in the
momentum space and is assumed to be a slow-varying smooth
function of $\bf k$.
Sometimes ${\cal M}(P,k)$ is a trace by which the on-shell spinors
$v(rP+k,s)$ and $\bar u(\bar r P-k,\bar s)$ are brought together
in the right order.
$\Gamma_{S S_Z} (P,k)$, up to second order in $k$, is given by
\begin{eqnarray}
\Gamma_{S S_Z}(P,k) &=& \sqrt{\frac{m_Q+m_q}{2m_Q m_q}} \; \sum_{s,\bar s} \;
\biggl\langle \frac{1}{2} s; \frac{1}{2} \bar s \biggr \vert
S S_Z  \biggr\rangle \;
v(r P +k,s) \; \bar u\left( \bar r P -k, \bar s \right)  \; , \nonumber \\
&\approx& \sqrt{m_Q +m_q} \left( \frac{r \overlay{/}{P} + \overlay{/}{k} -m_q}
{2 m_q} \right )\; \left (
\begin{array}{c}
\gamma^5 \\
- \overlay{/}{\epsilon}(P,S_Z)
\end{array}     \right ) \; \left( \frac{\bar r \overlay{/}{P} -\overlay{/}{k}
+m_Q}{2m_Q} \right ) \; ,
\label{Gamma}
\end{eqnarray}
where in the middle parenthesis $\gamma^5$ is for $S=S_Z=0$ (spin-singlet)
and $-\overlay{/}{\epsilon}(P,S_Z)$ is for $S=1$ (spin-triplet).

Since $k/M$ is a small quantity, we can expand ${\cal M}(P,k)$
around $k=0$ in a Taylor expansion:
\begin{equation}
\left. {\cal M}(P,k) = {\cal M}(P,0) + k_\alpha \;\frac{\partial {\cal M}(P,k)}
{\partial k_\alpha} \right |_{k=0} +
\left.
\frac{1}{2} k_\alpha k_\beta \; \frac{\partial^2 {\cal M}(P,k)}
{\partial k_\alpha k_\beta}  \right |_{k=0} + \cdots
\end{equation}
where the first, second, and third terms correspond to quantum numbers
$L=0, \, 1,$ and $2$ of the orbital angular momentum, and so forth.
Thus for the S-wave ($L=0$), P-wave ($L=1$), and D-wave ($L=2$)
states, the amplitude $A(P)$ will
depend on the radial wave-functions $R_S(0)$, $R_P^{'}(0)$, and
$R_D^{''}(0)$ through the following relations:
\begin{eqnarray}
\int \frac{ d^3 {\bf k}}{(2\pi)^3} \; \Psi_{00} ({\bf k} )
&=& \frac{R_S(0)}{\sqrt{4\pi}} \; , \nonumber \\
\int \frac{ d^3 {\bf k}}{(2\pi)^3} \; \Psi_{1L_Z} ({\bf k} ) k_\alpha
&=& - i \sqrt{\frac{3}{4\pi}} R^{'}_P(0) \epsilon_\alpha(P,L_Z) \; ,
\nonumber \\
\int \frac{d^3{\bf k}}{(2\pi)^3} \; \Psi_{2L_Z} ({\bf k} ) k_\alpha k_\beta
&=& \sqrt{\frac{15}{8\pi}} R^{''}_D(0) \epsilon_{\alpha\beta}(P,L_Z) \; ,
\end{eqnarray}
where $\epsilon_\alpha$ is the polarization vector for the
spin-1 particle, and
$\epsilon_{\alpha\beta}$ is the totally symmetric, traceless, and transverse
second rank polarization tensor for the spin-2 particle. In the D-wave case,
the amplitude becomes
\begin{equation}
\label{10}
A(P) =  \left.
\frac{1}{2} \sqrt{\frac{15}{8\pi}} \epsilon_{\alpha\beta}(P,J_Z) R_D^{''}(0)
\frac{\partial^2
{\cal M}(k)}{\partial k_\alpha \partial k_\beta}\right |_{k=0} \; ,
\end{equation}
for the spin-singlet case, where $J_Z=L_Z$, and
\begin{equation}
\label{11}
A(P)
= \left. \frac{1}{2} \sqrt{\frac{15}{8\pi}} R_D^{''}(0)
\Pi_{\alpha\beta\rho}^J (P,J_Z) \;
\frac{\partial^2 {\cal M}^\rho(k)}
{\partial k_\alpha \partial k_\beta}\right |_{k=0} \; ,
\label{rho}
\end{equation}
where
\begin{equation}
\Pi^J_{\alpha\beta\rho}(P,J_Z) =
\sum_{L_Z,S_Z}\;\epsilon_{\alpha\beta}(P,L_Z) \; \epsilon_\rho(P,S_Z)
\langle 2L_Z;1 S_Z | J J_Z \rangle \; ,
\end{equation}
for the spin-triplet case.
Using the appropriate Clebsch-Gordan coefficients, we have \cite{robin}
\begin{eqnarray}
\label{pi1}
\Pi_{\alpha\beta\rho}^{J=1}(P, J_Z) &=&
  - \sqrt{\frac{3}{20}} \; \left(
\frac{2}{3} {\cal P}_{\alpha\beta} \; \epsilon_\rho (P,J_Z)
-{\cal P}_{\alpha\rho}\;\epsilon_\beta(P,J_Z) - {\cal P}_{\beta\rho}
\;\epsilon_\alpha(P,J_Z) \right) \; , \\
\Pi_{\alpha\beta\rho}^{J=2} (P,J_Z) &=&
\frac{i}{M\sqrt{6}}\left( \epsilon_{\alpha\sigma}(P,J_Z)
\epsilon_{\tau\beta\rho
 \sigma'}\; P^\tau g^{\sigma\sigma'} + \epsilon_{\beta\sigma}(P,J_Z)
\epsilon_{\tau\alpha\rho\sigma'}\; P^\tau g^{\sigma\sigma'} \right) \; , \\
\Pi_{\alpha\beta\rho}^{J=3} (P,J_Z) &=& \epsilon_{\alpha\beta\rho} (P,J_Z) \; ,
\label{pi3}
\end{eqnarray}
where
\begin{equation}
{\cal P}_{\alpha\beta} = -g_{\alpha\beta} + \frac{P_\alpha P_\beta}{M^2}\;,
\end{equation}
and $\epsilon_{\alpha\beta\rho}(P,J_Z)$ is the totally symmetric, traceless,
and transverse spin-3 polarization tensor.
The polarization sums for $J=1, \, 2,$ and $3$ are given by the
following familiar expressions \cite{robin}
\begin{eqnarray}
\label{j=1}
\sum_{J_Z=-1}^1  \epsilon_\alpha(P,J_Z) \epsilon_\beta^*(P,J_Z) &=&
       {\cal P}_{\alpha\beta}  \; , \\
\label{j=2}
\sum_{J_Z=-2}^2  \epsilon_{\alpha\beta}(P,J_Z)
\epsilon_{\rho\sigma}^*(P,J_Z) &=&
\frac{1}{2}\left( {\cal P}_{\alpha\rho} {\cal P}_{\beta\sigma}
+ {\cal P}_{\alpha\sigma} {\cal P}_{\beta\rho}
\right ) -\frac{1}{3} {\cal P}_{\alpha\beta} {\cal P}_{\rho\sigma} \; , \\
\label{j=3}
\sum_{J_Z=-3}^3  \epsilon_{\alpha\beta\gamma}(P,J_Z)
		 \epsilon_{\rho\sigma\eta}^*(P,J_Z) &=&
\frac{1}{6} \biggr({\cal P}_{\alpha\rho} {\cal P}_{\beta\sigma}
{\cal P}_{\gamma\eta} + {\cal P}_{\alpha\rho} {\cal P}_{\beta\eta}
{\cal P}_{\gamma\sigma} + {\cal P}_{\alpha\sigma} {\cal P}_{\beta\rho}
{\cal P}_{\gamma\eta}
\nonumber \\
&& \quad         + {\cal P}_{\alpha\sigma} {\cal P}_{\beta\eta}
{\cal P}_{\gamma\rho} + {\cal P}_{\alpha\eta} {\cal P}_{\beta\sigma}
{\cal P}_{\gamma\rho} + {\cal P}_{\alpha\eta} {\cal P}_{\beta\rho}
{\cal P}_{\gamma\sigma} \biggr )
\nonumber \\
&& -\frac{1}{15} \biggr({\cal P}_{\alpha\beta} {\cal P}_{\gamma\eta}
{\cal P}_{\rho\sigma} + {\cal P}_{\alpha\beta} {\cal P}_{\gamma\sigma}
{\cal P}_{\rho\eta} + {\cal P}_{\alpha\beta} {\cal P}_{\gamma\rho}
{\cal P}_{\sigma\eta}
\nonumber \\
&& \quad              + {\cal P}_{\alpha\gamma} {\cal P}_{\beta\eta}
{\cal P}_{\rho\sigma} + {\cal P}_{\alpha\gamma} {\cal P}_{\beta\sigma}
{\cal P}_{\rho\eta} + {\cal P}_{\alpha\gamma} {\cal P}_{\beta\rho}
{\cal P}_{\sigma\eta}
\nonumber \\
&& \quad              + {\cal P}_{\beta\gamma} {\cal P}_{\alpha\eta}
{\cal P}_{\rho\sigma} + {\cal P}_{\beta\gamma} {\cal P}_{\alpha\sigma}
{\cal P}_{\rho\eta} + {\cal P}_{\beta\gamma} {\cal P}_{\alpha\rho}
{\cal P}_{\sigma\eta} \biggr) \; .
\end{eqnarray}

After writing down the formalism for the general production of D-wave
mesons, we shall specify the production mechanism.  The specific
production of D-wave $(Q\bar q)$ mesons that we are considering is by the
fragmentation of a heavy quark $Q$:
\begin{equation}\
\label{5}
Q^*(q) \to Q\bar q (P, \underline{1}) + q (p') \;,
\end{equation}
where the off-shell $Q^*$ is produced by a high energy source $\Gamma$.
The fragmentation function is then given by the formula
\cite{qswave1,bcswave}:
\begin{equation}
D(z) = \frac{1}{16\pi^2} \int ds \theta \left( s - \frac{M^2}{z} -
\frac{r^2M^2}{1-z} \right) \lim_{q_0 \to \infty}
\frac{\sum \vert A(P) \vert^2}{\sum \vert A_0 \vert^2} \; ,
\label{dz}
\end{equation}
where $\sum \vert A_0 \vert^2 = N_c {\rm Tr}(\Gamma \overline \Gamma
\overlay{/} q)$ is the tree-level amplitude squared to create an on-shell
$Q$ quark with the same 3-momentum ${\bf q}$,
and $\sum \vert A(P) \vert^2$ is the
amplitude squared for producing $(Q\bar q)+q$ from the same source $\Gamma$.
The general procedures to extract the fragmentation function
have been described in detail in Ref.\cite{qswave1,bcswave} and we shall be
brief in what follows.

\subsection{Unequal Mass Case}

We shall first derive the fragmentation functions for a heavy quark $Q$ into
an unequal mass D-wave $(Q\bar q)$ meson.
The fragmentation of $Q^*$ into a $Q\bar q$ meson is given by the
process in Eq.~(\ref{5}),
of which the leading Feynman diagram is given in Fig.~\ref{fig3}.
In reality, the above fragmentation process applies to
$b^*\to \overline{\delta^{bc}}+c$
($b^*\to \overline{\delta^{bc}_J}+c$)
and also its charge conjugate
$\bar b^* \to \delta^{bc} + \bar c$
($\bar b^* \to \delta^{bc}_J + \bar c$).
The amplitudes that are presented in the following are
for $Q^* \to (Q\bar q) + q$, while the amplitudes for the conjugate process
can be obtained by complex conjugation.  However, the final expressions
for the fragmentation functions are valid for both $Q^* \to (Q\bar q) +q$ and
$\overline{Q^*} \to (\overline Q q) + \bar q$ fragmentation processes.

The amplitude $A(P)$ for producing the spin-singlet D-wave state is
given in Eq.~(\ref{10}) with $\frac{\partial^2 {\cal M}}
{\partial k_\alpha \partial k_\beta}\biggr\vert_{k=0}$ given by
\begin{equation}
\label{M2}
\frac{\partial^2 {\cal M}}
{\partial k_\alpha \partial k_\beta}\biggr\vert_{k=0}
= N_{ij} \Delta_F \bar u (p^\prime) V^{\alpha\beta} \gamma_5 \Gamma \; ,
\end{equation}
where
\begin{eqnarray}
N_{ij} &=& g_s^2 C_F \frac{\delta_{ij}}{\sqrt{3}}
\frac{1}{4 r \bar r M {\sqrt M}}  \;\;\;\;\;\;
{\rm with} \; \; \; \; C_F \, = \, \frac{4}{3} \; ,
\\
V^{\alpha\beta} &=& V_1^{\alpha\beta} + (1 - 2r) V_2^{\alpha\beta} \; ,
\end{eqnarray}
and
\begin{eqnarray}
V_1^{\alpha\beta} & = &   \frac{32\bar r}{r^2} M \Delta_F^3 q^\alpha q^\beta
(\overlay{/}{P}+2M)(\overlay{/}{q}-\bar r M)
+\frac{16\bar r}{r^2}M \Delta_F^2 A_F
\;q^\alpha q^\beta\;\overlay{/}{n}\; (\overlay{/}{P}-M)
\nonumber\\
&+&\frac{2}{r^2} \Delta_F A_F \overlay{/}{n} \;
(q^\alpha \gamma^\beta + q^\beta \gamma^\alpha ) \overlay{/}{P}
+ \frac{4\bar r}{r} M \Delta_F A_F^2
(q^\alpha n^\beta + q^\beta n^\alpha)\; \overlay{/}{n} \;
(\overlay{/}{P}-M)
\nonumber \\
&+&\frac{1}{r} A_F^2 \;\overlay{/}{n} \;
(n^\alpha \gamma^\beta + n^\beta \gamma^\alpha) \overlay{/}{P}
+4 \bar r M A_F^3 n^\alpha n^\beta \;\overlay{/}{n} \;(\overlay{/}{P} - M)
\; ,
\\
V_2^{\alpha\beta} & = & - \frac{4}{r^2} M \Delta_F^2
(\gamma^\alpha q^\beta + \gamma^\beta q^\alpha) (\overlay{/}{q} - \bar r M)
- \frac{2}{r^2}M \Delta_F A_F \overlay{/}{n}
(\gamma^\alpha q^\beta + \gamma^\beta q^\alpha)
\nonumber \\
&-& \frac{1}{r} M A_F^2 \overlay{/}{n}
(\gamma^\alpha n^\beta + \gamma^\beta n^\alpha) \; ,
\label{vertex1}
\end{eqnarray}
with
\begin{eqnarray}
\Delta_F &=& (s - \bar r ^2 M^2)^{-1} \; ,
\\
A_F &=& \left( n \cdot ( q - \bar r P) \right)^{-1} \; .
\end{eqnarray}
In the above formulas, $s=q^2$ and we have chosen the axial gauge associated
with the 4-vector $n^\mu=(1,0,0,-1)$.  In this gauge, figure~\ref{fig3}
is the only leading Feynman diagram where factorization
between the short distance process and fragmentation becomes manifest!
We square the amplitude $A(P)$ and sum over colors and
helicities of $q$ and $(Q\bar q)$, and put them into Eq.~(\ref{dz}).
The heavy quark fragmentation function for the $^1D_2$ state thus obtained is
given by
\begin{eqnarray}
D_{Q \to Q\bar q(^1D_2)} (z) &=& \frac{\alpha_s^2(2rM)
|R^{''}_D(0)|^2}{324 \pi M^7}\;
\frac{1}{r^6\bar r^2} \; \frac{z(1-z)^2}{(1- \bar r z)^{10}} \biggr\{
60 +60(-7+8r-4r^2) z \nonumber \\
&+& 5(267 -600r +580r^2 -160r^3 +64r^4)z^2 \nonumber \\
&+& 10(-255 +825r -1130r^2 +596r^3 -88r^4 -64r^5 )z^3 \nonumber \\
&+&  (3225 -13050r +22585 r^2 -16872 r^3 +5036r^4 -576r^5 +1152 r^6) z^4
\nonumber\\
&-& 4 \bar r(690 -2535r +4200r^2 -2279 r^3 +968 r^4 -524 r^5 -160 r^6 )z^5
\nonumber \\
&+& \bar r^2 (1545-4890r +8525r^2 -2176r^3 +3076r^4 -320r^5 +320r^6)z^6
\nonumber \\
&-& 2 \bar r^3 (255-660r +1475r^2 +174r^3 +740 r^4 +120r^5)z^7 \nonumber\\
&+& 5\bar r^4 (15-30r+107r^2+56r^3 +80r^4)z^8 \biggr \} \; .
\end{eqnarray}
For the spin-triplet case, the amplitude $A(P)$ is given in Eq.~(\ref{11})
with $\frac{\partial^2 {\cal M}^\rho}
{\partial k_\alpha \partial k_\beta}\biggr\vert_{k=0}$ given by
\begin{equation}
\label{M3}
\frac{\partial^2 {\cal M}^\rho(k)}
{\partial k_\alpha \partial k_\beta}\biggr\vert_{k=0}
= N_{ij} \Delta_F \bar u (p^\prime)
V^{\alpha\beta\rho} \Gamma \; ,
\end{equation}
where
\begin{equation}
V^{\alpha\beta\rho} = V_1^{\alpha\beta\rho} +
(1 - 2r) V_2^{\alpha\beta\rho} \; ,
\end{equation}
and
\begin{eqnarray}
V_1^{\alpha\beta\rho} & = &
\frac{4}{r} \Delta_F \; \biggl[ (\gamma^\beta g^{\rho\alpha} + \gamma^\alpha
  g^{\rho\beta} )
- \frac{1}{r} \Delta_F\; \biggl( (q^\alpha (\gamma^\beta \gamma^\rho
-2 \bar r g^{\rho\beta})+q^\beta (\gamma^\alpha \gamma^\rho -
2 \bar r g^{\rho\alpha})
)\; \overlay{/}{P}  \nonumber \\
&& \qquad  + 2 M\; (q^\alpha g^{\rho\beta}
+q^\beta g^{\rho\alpha}) \biggr)
+\frac{8\bar r}{r} M^2 \Delta_F^2 q^\alpha q^\beta \; \gamma^\rho \biggr]
(\overlay{/}{q} + \bar r M)
\nonumber \\
&+&  \frac{1}{r} A_F \; \overlay{/}{n}
\biggr[ 2(\gamma^\alpha g^{\rho\beta} + \gamma^\beta g^{\rho\alpha} )
\nonumber \\
&& \qquad + A_F \biggr (
  n^\alpha( (\gamma^\beta \gamma^\rho - 2 r g^{\rho\beta})\overlay{/}{P}
             +2rMg^{\rho\beta} )  +
  n^\beta( (\gamma^\alpha \gamma^\rho - 2 r g^{\rho\alpha})\overlay{/}{P}
             +2rMg^{\rho\alpha} )  \biggr)
\nonumber \\
&& \qquad + \frac{2}{r} \Delta_F \biggr(
 q^\alpha ( (\gamma^\beta\gamma^\rho - 2rg^{\rho\beta}) \overlay{/}{P}
              +2rMg^{\rho\beta} )
+q^\beta ( (\gamma^\alpha\gamma^\rho - 2rg^{\rho\alpha})
\overlay{/}{P} +2rMg^{\rho\alpha} )  \biggr ) \biggr ]
\nonumber  \\
&-& 4 \bar r M A_F \; \left[
\frac{4}{r^2} \; q^\alpha q^\beta \Delta_F^2
+\frac{1}{r} \Delta_F A_F \;(q^\alpha n^\beta + q^\beta n^\alpha)
\;  + A_F^2 \; n^\alpha n^\beta \right]\;
   \overlay{/}{n} (\overlay{/}{P}-M) \gamma^\rho  \; , \\
V_2^{\alpha\beta\rho}  &=&
\frac{M}{r} A_F \; \overlay{/}{n} \biggl[
\frac{2}{r} \Delta_F \left( q^\alpha \gamma^\beta + q^\beta \gamma^\alpha
\right ) +
A_F \left( n^\alpha \gamma^\beta + n^\beta \gamma^\alpha \right ) \biggr]
\gamma^\rho  \; .
\label{vertex2}
\end{eqnarray}
We then square the amplitude and sum over the colors and helicities of
$q$ and $(Q\bar q)$ using the polarization sum formulas in Eqs.~(\ref{j=1}) --
(\ref{j=3}).
The  fragmentation functions thus obtained are
$$
D_{Q \to Q\bar q(^3D_1)} (z) = \frac{\alpha_s^2(2rM) |R^{''}_D(0)|^2}
{3240 \pi M^7}\;
\frac{1}{r^6\bar r^2} \; \frac{z(1-z)^2}{(1-\bar r z)^{10}}
\qquad \qquad \qquad \qquad \qquad \qquad
\qquad
$$
\begin{eqnarray}
&\times & \biggr \{  60(1+4r)^2
- 60(1+4r)(7+36r -28r^2) z \nonumber \\
&+& 5(261 + 2652r + 6128r^2 - 12576r^3 + 7360r^4)
z^2 \nonumber \\
&+&10 (-237 - 2589r - 8704r^2 + 27328r^3 - 28608r^4 + 11360r^5)z^3 \nonumber\\
&+& (2775 + 31350r + 153935r^2 - 637872r^3 + 897816r^4 - 611056r^5 +
     183552r^6)z^4 \nonumber \\
&-& 4\bar r (540+6585r +48660r^2 -162629r^3 +199928r^4 -118084r^5
+28400r^6)z^5 \nonumber\\
&+&  \bar r^2  (1095 + 13830 r +137935r^2 - 376916r^3 + 385696r^4 - 192320r^5
+ 36800r^6 )z^6 \nonumber \\
&-&2\bar r^3  (165 +2100r +26395r^2 -55836r^3 +48360r^4 -26720r^5 +
     3360r^6 )z^7 \nonumber \\
&+& 5 \bar r^4   (9 + 114r + 1717r^2 - 2424r^3 + 2476r^4 - 1872r^5 +
     192r^6 )z^8  \biggr \} \; ,
\end{eqnarray}
\begin{eqnarray}
D_{Q \to Q\bar q(^3D_2)} (z)
&=& \frac{\alpha_s^2(2rM) |R^{''}_D(0)|^2}{648 \pi M^7}\;
\frac{1}{r^6\bar r^2} \; \frac{z(1-z)^2}{(1-\bar r z)^{10}}
\biggr \{  180 +180(-7+4r)z  \nonumber \\
&+& 5(777-852r +464r^2 -64r^3 +128r^4)z^2 \nonumber \\
&+& 10( -693 +1087r -1126 r^2 +624r^3 -112r^4 -128r^5) z^3 \nonumber \\
&+& (7875 -15650r +23295r^2 -22564r^3 +9112r^4 +128r^5 +2304r^6)z^4 \nonumber\\
&-& 4\bar r (1470 -2005r + 4740r^2 -4613r^3 +2296r^4 -488r^5 -320r^6)z^5
\nonumber \\
&+& \bar r^2(2835-2050r +12035r^2 -8332r^3 +5952 r^4 -960r^5 +640r^6)z^6
\nonumber \\
&-& 2\bar r^3 (405 -40r +2565r^2 -602r^3 +1000r^4 -320r^5) z^7 \nonumber \\
&+& 5\bar r^4 (21+10r +205r^2 +52r^3 +84r^4)z^8 \biggr \} \; , \\
D_{Q \to Q\bar q(^3D_3)} (z)
&=& \frac{2\alpha_s^2(2rM) |R^{''}_D(0)|^2}{405 \pi M^7}\;
\frac{1}{r^6} \; \frac{z(1-z)^2}{(1-\bar r z)^{10}}
\biggr \{  90 + 90(-7+2r)z  \nonumber \\
&+& 5(399-174r+200r^2) z^2 +10(-378 +193r -437r^2 +70r^3)z^3 \nonumber \\
&+& (4725 -2900r +8390r^2 -1918r^3 +2268r^4 )z^4 \nonumber \\
&+& 2(-1995 +1700r -4930r^2 +1617r^3 -2387r^4 +350r^5 )z^5 \nonumber \\
&+& (2205 -2830r +7940r^2 -5194r^3 +4914r^4 -930r^5 +1000r^6) z^6 \nonumber \\
&-& 2\bar r (360-325r +1680r^2 -749r^3 +1540r^4 +295r^5 +90r^6) z^7 \nonumber\\
&+& 5\bar r^2 (21-14r +133r^2 -56r^3 +189r^4 -18r^5 +18r^6)z^8 \biggr \} \; .
\end{eqnarray}
All the above fragmentation functions are valid for the fragmentation of
$\bar b \to \delta^{bc}$ and $\delta^{bc}_J$, as well as for
$b \to \overline{\delta^{bc}}$ and $\overline{\delta^{bc}_J}$, where $J=1,2,3$.

For a given principal quantum number $n$ the $^1D_2$ and $^3D_2$ states
constructed in the LS coupling scheme are mixed in general to form the
physical states $2^+$ and $2^{+'}$ defined as
\begin{equation}
\label{mix}
\begin{array}{rcl}
|2^{+'} \rangle &=& \cos\theta |^1D_2 \rangle -
 \sin\theta |^3D_2 \rangle \; ,\\
|2^{+} \rangle &=& \sin\theta |^1D_2 \rangle + \cos\theta |^3D_2 \rangle \;,
\end{array}
\end{equation}
where $\theta$ is the mixing angle.  Thus, the
fragmentation functions into the physical states $2^+$ and $2^{+'}$ are
\begin{equation}
\label{mix2}
\begin{array}{rcl}
D_{Q \to Q\bar q(2^{+'})}(z) &=& \cos^2\theta D_{Q \to Q\bar q(^1D_2)}(z) +
\sin^2\theta D_{Q \to Q\bar q(^3D_2)}(z) -\sin\theta\cos\theta
D_{\rm mix}(z) \; ,\\
D_{Q\to Q\bar q (2^+)}(z) &=& \sin^2\theta D_{Q \to Q\bar q(^1D_2)}(z) +
\cos^2\theta D_{Q\to Q\bar q(^3D_2)}(z) +\sin\theta\cos\theta
D_{\rm mix}(z) \; .
\end{array}
\end{equation}
Therefore, we also need to calculate the fragmentation function
$D_{\rm mix}(z) $.  We note that in the case of charmonium
it is the charge-conjugation quantum number $C$ that prevents the mixing
of $c\bar c(^1D_2)$ and $c\bar c(^3D_2)$ states, because $c\bar c (^1D_2)$
has $C=+1$ while $c\bar c (^3D_2)$ has $C=-1$.  In the unequal-mass case
the meson does not have this quantum number $C$ to prevent the mixing.
The mixing fragmentation function is obtained by calculating the
interference term $A(^1D_2) A^*(^3D_2) + A^*(^1D_2) A(^3D_2)$
and sum over the colors and helicities of the particles in the final state as
before. The mixing fragmentation function thus obtained is
\begin{eqnarray}
D_{\rm mix}(z) &=& - \frac{5 \alpha_s^2(2rM) |R^{''}_D(0)|^2}
{54\sqrt{6}\pi M^7}\; \frac{1}{r^6\bar r^2}\; \frac{z(1-z)^2}{(1-\bar r z)^8}
\biggr \{ 12 +12(-5+4r -2r^2)z \nonumber \\
&+& (129 -198r +156r^2 +64r^3) z^2 +4(-39 +85r -76r^2 +8r^3 -36r^4)z^3
\nonumber \\
&+& 2(57-156r +133r^2 -130r^3 +198r^4 +48r^5)z^4 \nonumber \\
&-& 4\bar r(12-27r +4r^2 -51r^3 -2r^4 -14r^5)z^5 \nonumber \\
&+& \bar r^2 (9-16r -11r^2 -62r^3 -4r^4) z^6 \biggr \} \; .
\end{eqnarray}

The fragmentation probabilities for these states can be obtained by integrating
the fragmentation functions over $z$,
\begin{eqnarray}
\label{prob1}
\int_0^1 D_{Q\to Q\bar q(^1D_2)} (z) dz &=& \frac{10 N}{21 r^7 \bar r^{10}}
\; \biggr[ \bar r(793 +3712r +42160r^2 +123220r^3 +253315r^4 \nonumber \\
&+& 126622r^5  +9778r^6 +1180r^7 -80r^8 ) \nonumber\\
&+&  315 r (11+40r +280r^2 +580r^3 +701r^4 +152r^5 +16r^6) \log r \biggr],
\nonumber \\
&& \\
\int_0^1 D_{Q\to Q\bar q(^3D_1)} (z) dz &=& \frac{N}{7 r^7 \bar r^{10}}\;
\biggr[\bar r(209+457r +28224r^2+479388r^3 +1136191r^4 \nonumber \\
&-& 358043r^5 -430066r^6 -113308r^7 +12528r^8) \nonumber \\
&+& 105r (5+60r +1772r^2 +8644r^3 +4759r^4 -5632r^5 -2108r^6
\nonumber \\
&-& 368r^7 +64 r^8) \log r \biggr] \; , \\
\int_0^1 D_{Q\to Q\bar q(^3D_2)} (z) dz &=& \frac{5N}{21 r^7 \bar r^{10}}\;
\biggr[\bar r(1127 -700r +67880r^2 +320552r^3 +393317r^4 \nonumber \\
&+& 191990r^5 +118742r^6 +4712r^7 -160r^8 ) \nonumber \\
&+& 315 r ( 9 +44r +604r^2 +1260r^3 +879r^4 +532r^5 +156r^6)
\log r \biggr] \; , \nonumber \\
&& \; \\
\int_0^1 D_{Q\to Q\bar q(^3D_3)} (z) dz &=& \frac{8N}{21 r^7 \bar r^{10}}\;
\biggr[ \bar r(2191 +4828r +100786r^2 +160027r^3 +288799r^4 \nonumber \\
&+& 428008r^5 -63854r^6 +87913r^7 +14422r^8 ) \nonumber \\
&+& 630 r (15 +20r +343r^2 +126 r^3 +861r^4 +112r^5 +63r^6 \nonumber\\
&+& 78r^7  +6r^8 ) \log r \biggr]  \; ,
\label{prob2} \\
\int_0^1 D_{\rm mix} (z) dz &=& -\frac{10\sqrt{6}N}
{7 r^7 \bar r^{10}}\;
\biggr[\bar r(153 +358r +3214 r^2 -13796r^3 -15021 r^4 \nonumber \\
&+& 7176r^5 +9696r^6 +240r^7 ) \nonumber \\
&+& 105 r ( 5+10r -20r^2 -180r^3 -37r^4 +110r^5 +36r^6 ) \log r \biggr] \; ,
\end{eqnarray}
where $N=\alpha_s^2(2rM) |R^{''}_D(0)|^2/(3240 \pi M^7)$.
We note that the running scale used in the strong coupling constant
$\alpha_s$ is set at $2rM$, which is the minimal virtuality of the exchanged
gluon \cite{qswave1}.

\subsection{Equal Mass Case}

Heavy quark fragmentation functions for the D-wave quarkonium can be obtained
simply by setting $r=1/2$ and $\theta = 0$ in the previous formulas for the
unequal mass case.   For convenience we also present their explicit
formulas in the following.
\begin{eqnarray}
D_{Q \to \delta^Q}(z) &=& \frac{8\alpha_s^2(2m_Q) |R_D^{''}(0)|^2}{81\pi m_Q^7}
\frac{z(1-z)^2}{(2-z)^{10}} \left( 3840 -15360z + 30720 z^2 - 37120 z^3 \right.
\nonumber\\
&+ & \left.  35328z^4 -29344z^5 +18344z^6 - 5848z^7 +775 z^8 \right ) \; ,
\end{eqnarray}
\begin{eqnarray}
D_{Q\to \delta^Q_1}(z) &=& \frac{8 \alpha_s^2(2m_Q)
|R_D^{''}(0)|^2}{405\pi m_Q^7} \;
\frac{z(1-z)^2}{(2-z)^{10}}\; \left( 17280 -103680z +321120z^2 -551840z^3
\right. \nonumber\\
&+& \left. \, 546744z^4 -314752z^5 + 112402z^6 -24594z^7 + 2915 z^8 \right)
\; ,\\
D_{Q\to \delta^Q_2}(z) &=& \frac{16 \alpha_s^2(2m_Q)
|R_D^{''}(0)|^2}{81\pi m_Q^7} \;
\frac{z(1-z)^2}{(2-z)^{10}}\; \left( 2880 -14400z +37360z^2 -58240z^3
\right. \nonumber\\
&+& \left. \, 58604z^4 -38372z^5 +16517z^6 -4014z^7 +445z^8 \right ) \; , \\
D_{Q\to \delta^Q_3}(z) &=& \frac{8 \alpha_s^2(2m_Q)
|R_D^{''}(0)|^2}{405\pi m_Q^7} \;
\frac{z(1-z)^2}{(2-z)^{10}}\; \left( 11520 -69120z +231680z^2 -488960z^3
\right. \nonumber\\
&+& \left.  \, 675136z^4 -592288z^5 +309688z^6 -80736z^7 + 8285z^8 \right ) \;.
\end{eqnarray}
We can also obtain the fragmentation probabilities by integrating over $z$:
\begin{eqnarray}
\int_0^1 dz D_{Q\to \delta^Q}(z) & = &
\frac{8\alpha_s^2(2m_Q) |R_D^{''}(0)|^2}{81\pi m_Q^7}
\left( \frac{776677}{21} - 53355 \, {\rm log} \, 2 \right) \; , \\
\int_0^1 dz D_{Q\to \delta^Q_1}(z) & = &
\frac{2\alpha_s^2(2m_Q) |R_D^{''}(0)|^2}{405\pi m_Q^7}
\left( 547127 - 789300\, {\rm log} \, 2 \right) \; , \\
\int_0^1 dz D_{Q\to \delta^Q_2}(z) & = &
\frac{16\alpha_s^2(2m_Q) |R_D^{''}(0)|^2}{81\pi m_Q^7}
\left( \frac{2889265}{168} - 24810 \, {\rm log} \, 2 \right) \; , \\
\int_0^1 dz D_{Q\to \delta^Q_3}(z) & = &
\frac{8\alpha_s^2(2m_Q) |R_D^{''}(0)|^2}{405\pi m_Q^7}
\left( \frac{5182432}{21} - 356025 \, {\rm log} \, 2 \right) \; .
\end{eqnarray}

\section{Discussions}

We have seen that the covariant formalism used to calculate the
D-wave fragmentation functions is quite cumbersome and tedious. The
complicated  results are therefore needed to be cross-checked by other means.
To be sure we have recalculated the D-wave fragmentation functions for
the equal mass case from scratch and we did end up with the same results
as what we have in the above for $r=1/2$.
Two additional nontrivial checks are performed by using
the Braaten-Levin spin-counting rule and the heavy-quark spin-counting rule
which we will discuss in turns.

\subsection{Braaten-Levin Spin-counting Rule}

The fragmentation function $D_{i \to H}(z)$ for a parton $i$ splitting into a
hadron $H$ is related to the distribution function $f_{i/H}(x)$
of finding the parton $i$ inside the hadron $H$
by the analytic continuation \cite{drelletal,lipatovetal}
\begin{equation}
\label{79}
f_{i/H}(x) = x D_{i \to H}\left(\frac{1}{x} \right ) \; .
\end{equation}
The results of perturbative fragmentation functions
allow us to study the perturbative tail of the distribution functions of
the heavy quark inside the heavy mesons as well.
{}From our explicit calculations, we see that $f_{i/H}(x)$ has a pole
located at $x=\bar r$.
The pole is cut off by non-perturbative effects related to the formation of
bound states of the $(Q\bar q)$ pair.
This pole is of order 6, 8 and 10 for  S-, P-, and D-wave states,
respectively. In the general case of L-waves
we expect this pole is of order $6+2L$. Therefore, we can expand $f(x)$ as a
Laurent series,
\begin{equation}
f(x) = {a_{n}(r) \over (x - \bar r)^n} \; +
{a_{n-1}(r) \over (x - \bar r)^{n-1}} \; + \;
{\rm less \; singular \; terms} \; ,
\label{blexp}
\end{equation}
with $n=6+2L$ for the general L-waves.

The Braaten-Levin rule states that the leading $r$-dependent
coefficients ${a_n(r)}$ satisfy the simple spin-counting.
Specifically, we have
\begin{equation}
a_6(^1S_0):a_6(^3S_1) \; = \; 1:3 \;\;\; \mbox{for S-waves},
\end{equation}
\begin{equation}
a_8(^1P_1):a_8(^3P_0):a_8(^3P_1):a_8(^3P_2) \; = \;
3:1:3:5 \;\;\;\; \mbox{for P-waves},
\end{equation}
\begin{equation}
a_{10}(^1D_2):a_{10}(^3D_1):a_{10}(^3D_2):a_{10}(^3D_3) \; = \;
5:3:5:7 \;\; \mbox{for D-waves},
\end{equation}
and so on.

The applicability of these counting rules has been demonstrated
for the S- and P-wave cases in Ref.~\cite{qpwave1}.  For the D-wave case
we can expand the distribution functions, which are obtained from the
D-wave fragmentation functions by Eq.~(\ref{79}),  around $x=\bar r$
and obtain the first two terms in the Laurent series:
\begin{eqnarray}
f_{Q / \delta^{Q \bar q}} (x)
& = &  \frac{30720 N (r\bar r)^2}{(x-\bar r)^{10}} +
\frac{320 N (576 r-275) r \bar r}{(x-\bar r)^9} +
\cdots \; ,\\
f_{Q / \delta^{Q \bar q}_1} (x)
& = &  \frac{18432 N (r\bar r)^2}{(x-\bar r)^{10}} +
\frac{192 N (576 r-275) r \bar r}{(x-\bar r)^9} +
\cdots \; , \\
f_{Q / \delta^{Q \bar q}_2} (x)
& = &  \frac{30720 N (r\bar r)^2}{(x-\bar r)^{10}} +
\frac{320 N (576 r-275) r \bar r}{(x-\bar r)^9}
+ \cdots \; , \\
f_{Q / \delta^{Q \bar q}_3} (x)
& = &  \frac{43008 N (r\bar r)^2}{(x-\bar r)^{10}} +
\frac{448 N (576 r-275) r \bar r}{(x-\bar r)^9} +
\cdots \; ,
\end{eqnarray}
where $N=\alpha_s^2 |R^{''}_D(0)|^2/(3240 \pi M^7)$.  Therefore, the leading
coefficients $a_{10}$'s indeed satisfy the spin-counting
ratio $5:3:5:7$.  Actually, we
found that not only the first term but also the second term in the Laurent
series of $f(x)$ obey the Braaten-Levin spin-counting rules,
while the third term does not. We conjecture that the Braaten-Levin
spin-counting rule can be applied to the first two coefficients
in the expansion (\ref{blexp}) for all L-wave (L = 0, 1, 2, ...) cases.

\subsection{Heavy-Light Limit}

Hadrons containing a single heavy quark exhibit heavy quark symmetry  in
the limit $m_Q/\Lambda_{\rm QCD} \to \infty$.
In the limit of $m_Q / \Lambda_{QCD} \to \infty$, both
the heavy quark spin $\vec S_Q$
and the total spin $\vec J$ of a heavy hadron containing a
single heavy quark $Q$ become good quantum numbers.
This implies that in the spectroscopy of the hadron containing a
single heavy quark $Q$, the angular momentum of the light degrees of freedom
$\vec J_l = \vec J - \vec S_Q$ is
also a good quantum number.  We refer collectively to
all the degrees of freedom in the heavy-light hadron
other than the heavy quark as the light degrees of freedom.
For heavy-light $(Q\bar q)$ mesons, $\vec J_l = \vec S_q + \vec L$
where $\vec S_q$ is the spin of the light quark $q$
and $\vec L$ is the orbital angular momentum.
Thus, hadronic states can be labeled simultaneously by the eigenvalues
$j$ and $j_l$ of the total spin $\vec J$ and the angular momentum of the
light degrees of freedom $\vec J_l$, respectively.
In general \cite{spectrum}, the spectrum of hadrons containing
a single heavy quark has, for each $j_l$, a degenerate doublet with
total spins $j_+ = j_l + 1/2$ and $j_- =  j_l - 1/2$. (For the case of
$j_l=0$, the total spin must be $1/2$.) For D-wave heavy-light mesons,
$j_l$ can either be 3/2 or 5/2.
Thus $(j_-,j_+)=(1,2)$ and (2,3) for $j_l$ = 3/2 and 5/2, respectively.
As a result, we expect to have two distinct doublets $(^3D_1,2^{+'})$ and
$(2^{+},^3D_3)$ in the limit of $m_Q/m_q \rightarrow \infty$, {\it i.e.}
$r \to 0$.   In this limit,
the mixing coefficients in Eq.~(\ref{mix})
can be determined by the Clebsch-Gordan coefficients
in the tensor product of a spin $1/2$ state and a spin $2$ state
with the following result:
\begin{eqnarray}
|2^{+'} \rangle & = &   \sqrt{\frac{2}{5}} |^1D_2 \rangle
+  \sqrt{\frac{3}{5}} |^3D_2 \rangle \; , \\
|2^{+} \rangle &  =  &
 -\sqrt{\frac{3}{5}} |^1D_2 \rangle + \sqrt{\frac{2}{5}} |^3D_2 \rangle  \; ,
\label{mixhqlimit}
\end{eqnarray}
{\it i.e.},  we are transforming  the states $^1D_2$ and $^3D_2$ in the LS
coupling scheme to the states $2^{+}$ and $2^{+'}$ in the $jj$ coupling scheme.

In their discussions of the heavy quark fragmentation functions within the
context of Heavy Quark Effective Theory, Jaffe and Randall \cite{jaffe}
showed that fragmentation functions can have a $1/m_Q$ expansion if expanded
in terms of a more natural variable
\begin{equation}
y = {{1 \over z} - \bar r \over r} \; ,
\end{equation}
rather than the usual fragmentation variable $z$, and
the heavy-quark mass expansion is given as a power series in $r$,
\begin{equation}
D(y) = {1 \over r}a(y) + b(y) + {\cal O}(r) \; ,
\end{equation}
where $a(y), b(y)$, {\it etc.} are functions of the variable $y$.
The leading term $a(y)$ is constrained by the heavy quark spin-flavor
symmetry while all the higher order terms contain spin-flavor
symmetry breaking effects. One can recast our results for the D-wave
fragmentation functions derived in Sec.III in the above form, by
carefully expanding the powers of $r$ and $(1 - \bar r z)$.
The leading $1/r$ terms of the fragmentation functions are then given by
\begin{eqnarray}
D_{Q \to \delta^{Q \bar q}}(y)
&\to & \frac{10N'(y-1)^2}{ry^{10}}\; (3072 -2656 y +2056 y^2
      -912y^3 +660y^4 -120y^5 +75 y^6 )  \, , \nonumber \\
D_{Q \to \delta^{Q \bar q}_1}(y)
&\to & \frac{3N'(y-1)^2}{ry^{10}}\; (6144 -5312 y + 1952 y^2
      -1824 y^3 +1020y^4 -120y^5 +15 y^6 ) \, , \nonumber \\
D_{Q \to \delta^{Q \bar q}_2}(y)
&\to & \frac{5N'(y-1)^2}{ry^{10}}\; (6144 -5312 y + 3392 y^2
      -1824 y^3 +1220y^4 -200y^5 +105 y^6 ), \nonumber \\
D_{Q \to \delta^{Q \bar q}_3}(y)
&\to & \frac{112N'(y-1)^2}{ry^{10}}\; (384 -332 y + 347 y^2
      -114 y^3 +95y^4 -20y^5 +15 y^6 ) \, , \nonumber \\
D_{\rm mix}(y)
&\to & - \frac{50\sqrt{6}N'(y-1)^2}{ry^{10}}\;  (144 y^2
+20y^4 -8 y^5 + 9y^6 ) \, , \nonumber \\
D_{Q \to 2^{+'}}(y)
&\to & \frac{5N'(y-1)^2}{ry^{10}} \; (6144 -5312 y +1952 y^2
    -1824 y^3 +1020 y^4 -120 y^5 +15 y^6 ) \, , \nonumber \\
D_{Q \to 2^{+}}(y)
&\to & \frac{80N'(y-1)^2}{ry^{10}}\; (384 -332y +347y^2
-114y^3      +95 y^4 -20 y^5 +15 y^6 )  \, , \nonumber
\end{eqnarray}
where $N'= \alpha_s^2 |R^{''}_D(0)|^2/(3240 \pi (rM)^7)$.  Thus,
at leading-order of $1/m_Q$,
we obtain the following spin-counting ratios
\begin{eqnarray}
\frac{D_{Q \to \delta^{Q \bar q}_1} (y)}{D_{Q \to 2^{+'}} (y)} & \to &
\frac{3}{5} \; , \\
\frac{D_{Q \to 2^{+}} (y) }{D_{Q \to \delta^{Q \bar q}_3} (y) } & \to &
\frac{5}{7} \;,
\end{eqnarray}
as expected from heavy quark spin symmetry.

\section{Conclusions}

The detections of the D-wave orbitally excited states of
heavy-heavy mesons are much more difficult than their ground states.
So far, none of D-wave quarkonium has been identified.  The basic
reasons are
(i) the production rates are small,
(ii) small branching ratios in subsequent decays, and
(iii) the very small efficiencies in subsequent levels of identification.
{}From the potential model calculations, D-wave charmonium are
likely to be above the $D\overline D$ threshold and hence
they will decay predominantly into a pair of D mesons.
In principle, D-wave quarkonium
can be identified partially by their pure leptonic decay
(e.g. $^3D_1 \to \mu^+\mu^-$), or
by their subsequent decays into the lower-lying states.
It is because of the subsequent
levels of decays that reduce substantially the branching ratios and
efficiencies in identifying these D-wave bound states. However, both
the branching ratios and efficiencies are uncertain at this stage.
In the following, we shall estimate the production rates of the
D-wave charmonium, bottomonium, and $(\bar bc)$ mesons based upon the
fragmentation functions obtained in Sec. III and IV.

The values of $|R^{''}_D(0)|^2$
for the $(c\bar c)$, $(b\bar b)$, and $(\bar bc)$ systems were
summarized nicely in Ref.\cite{quigg2} using different potential models.
We choose the QCD-motivated potential of Buchm\"{u}ller and Tye \cite{tye},
and the corresponding
values of $|R^{''}_D(0)|^2$ for the $(c\bar c)$, $(b\bar b)$,
and $(\bar bc)$ systems are 0.015, 0.637, and 0.055 GeV$^7$, respectively.
These values are for the first set of the D-wave states.
We note that the values of $|R^{''}_D(0)|^2$ obtained using the
Buchm\"{u}ller-Tye potential are the smallest ones
of the four potential models studied in Ref.\cite{quigg2}.

In the following analysis, we will adopt
the values $m_c=1.5$ GeV and $m_b=4.9$ GeV that were used in
the fit of the Buchm\"{u}ller-Tye potential to the bound state
spectra \cite{quigg2}.
We first evaluate the fragmentation probabilities for a $\bar b$ antiquark
into D-wave $(\bar bc)$ mesons by substituting $r=m_c/(m_b+m_c)=0.234$ into
Eqs.~(\ref{prob1}) -- (\ref{prob2}). Using $\alpha_s(2m_c)\approx 0.253$,
$|R^{''}_D(0)|^2=0.055 {\rm GeV}^7$ for the $(\bar bc)$ system, and
ignoring the possibly small mixing effects, we obtain
\begin{eqnarray}
\int_0^1 D_{\bar b \to \delta^{bc}}  (z) dz &=&  6.7 \times 10^{-6}  \; , \\
\int_0^1 D_{\bar b \to \delta^{bc}_1}(z) dz &=&  1.7 \times 10^{-6} \; ,  \\
\int_0^1 D_{\bar b \to \delta^{bc}_2}(z) dz &=&  6.5 \times 10^{-6} \; , \\
\int_0^1 D_{\bar b \to \delta^{bc}_3}(z) dz &=&  8.5 \times 10^{-6} \; .
\end{eqnarray}
Therefore, the total probability is about $2.3 \times 10^{-5}$.  We can
compare this value with the corresponding probabilities of the S-wave
and P-wave states.  Using the same $\alpha_s$ and the values of $|R_S(0)|^2$
and $|R^{'}_P(0)|^2$ calculated with the same Buchm\"{u}ller-Tye potential,
the total fragmentation probabilities of the 1S and 1P $(\bar b c)$
states are about $1\times 10^{-3}$ and $1.7 \times 10^{-4}$,
respectively \cite{qpwave1}.
These total fragmentation probabilities show a similar suppression
factor of order $v^2$ in going from 1S to 1P states and from
1P to 1D states, which are in accord with the velocity
counting rules in NRQCD \cite{bbl}.

Next, for the D-wave charmonium we use $\alpha_s(2m_c) \approx 0.253$ and
$|R^{''}_D(0)|^2=0.015 {\rm GeV}^7$,  and obtain
\begin{eqnarray}
\int_0^1 D_{c \to \delta^c}  (z) dz &=&  3.1 \times 10^{-6} \; , \\
\int_0^1 D_{c \to \delta^c_1}(z) dz &=&  2.3 \times 10^{-6} \; , \\
\int_0^1 D_{c \to \delta^c_2}(z) dz &=&  3.6 \times 10^{-6}  \; , \\
\int_0^1 D_{c \to \delta^c_3}(z) dz &=&  1.7 \times 10^{-6}  \; .
\end{eqnarray}
The total probability is about $1.1\times 10^{-5}$, to be compared with
the probabilities of $P(c\to J/\psi)\approx 1.8 \times 10^{-4}$ \cite{qswave1}
and $P(c\to h_c,\chi_c) \approx 8\times 10^{-5}$ \cite{qpwave1}
obtained using the same inputs.
With our inputs, the gluon fragmentation probability into the
$^1D_2$ charmonium \cite{gdwave}
is about $1.2 \times 10^{-6}$.
Similarly, using $\alpha_s(2m_b)\approx 0.174$ and
$|R^{''}_D(0)|^2=0.637 {\rm GeV}^7$ for the bottomonium, we obtain
\begin{eqnarray}
\int_0^1 D_{b \to \delta^b}  (z) dz &=&  1.6 \times 10^{-8} \; , \\
\int_0^1 D_{b \to \delta^b_1}(z) dz &=&  1.2 \times 10^{-8} \; , \\
\int_0^1 D_{b \to \delta^b_2}(z) dz &=&  1.8 \times 10^{-8} \; , \\
\int_0^1 D_{b \to \delta^b_3}(z) dz &=&  8.5 \times 10^{-9} \; .
\end{eqnarray}
The total probability for the D-wave bottomonium is very small
$\sim 5.4\times 10^{-8}$. This is to be compared with the probabilities of
$P(b\to \Upsilon)\approx 2 \times 10^{-5}$ \cite{qswave1}
and $P(b\to h_b, \chi_b) \approx 2.5 \times 10^{-6}$.
We observe that the relative probabilities for the S-wave to P-wave quarkonium
and for the P-wave to D-wave quarkonium are roughly
suppressed by a similar factor of $v^2$ as in the $\bar b c$ case.

At the Tevatron the $b$-quark production cross section is
of order $10 \mu b$ with $p_T \agt 6$ GeV, which implies about
$10^9$ $b$-quarks with an accumulated luminosity of 100 $pb^{-1}$.
Using the fragmentation probabilities calculated above,
we expect about $10^4$ D-wave $(\bar bc)$ mesons,
while there are only about 50 D-wave bottomonium.
At large $p_T$, charm quark production is very similar to
bottom quark production. Therefore,
the $c$-quark production cross section is also of order
10 $\mu b$ and there should be about $10^4$ D-wave charmonium for the
same luminosity.
Although the production rates of D-wave $(\bar bc)$ mesons and charmonium
are rather substantial, the small branching ratios and efficiencies of
identification in subsequent levels of decays render their detections
extremely difficult.  For example,
the combined branching ratio Br$({^1D_2} (c\bar c) \to {^1P_1} \gamma,
{^1P_1} \to J/\psi \pi^0, J/\psi \to \mu^+ \mu^-)$ has been
estimated to be about $10^{-4}$ \cite{gdwave}.
Even after the installation of the Main Injector when the yearly
luminosity can be boosted to 1 -- 3 $fb^{-1}$, the chance of detecting the
D-wave $(\bar bc)$ mesons or charmonium is still quite slim.
The future Large Hadron Collider (LHC) can produce $10^{12} - 10^{13}$
$b\bar b$ pairs and $c\bar c$ pairs per year running.
It can then produce about $10^7 - 10^8$ D-wave $(\bar b c)$ mesons
and charmonium.  After taking into account the branching ratios
and efficiencies, there should be a handful number of these D-wave states
that can be identified.
Nevertheless, detection of the D-wave bottomonium does not
seem to be feasible even at the LHC!
The production of $b\bar b$ and $c\bar c$ pairs at LEP,
LEPII, and other higher energy $e^+e^-$ or $\mu^+\mu^-$
colliders are much smaller than at the hadronic colliders.
Therefore, it is much more difficult to identify
the D-wave quarkonium or $(\bar bc)$ mesons
at the $e^+e^-$ or $\mu^+\mu^-$ facilities.

In conclusions, we have computed the heavy quark fragmentation functions
into D-wave heavy-heavy mesons containing two heavy quarks
to  leading order in strong coupling constant and to
leading order in the non-relativistic
expansion within the framework of factorization model of
Bodwin, Braaten, and Lepage.
The color-singlet contributions can be expressed
in terms of the NRQCD matrix elements
which are related to the second derivative of the non-relativistic
radial D-wave wave-function.
The color-octet contributions are expected to be quite small
in the heavy quark fragmentation and therefore
we have ignored them in the present analysis.
The fragmentation functions obtained in this work should be useful
in the production of the D-wave charmonium, bottomonium, and $(\bar bc)$
mesons at the future Large Hadronic Collider.

\section*{Acknowledgment}
This work was supported in part by the United States Department of
Energy under Grant Numbers DE-FG03-93ER40757 and DE-FG03-91ER40674.


\section*{Figures Caption}
\begin{enumerate}

\item \label{fig1}
Feynman diagram for $Q^*(q) \to Q \overline Q  (P, \underline 1 \; {\rm or} \;
\underline 8) + Q(p')$.

\item \label{fig2}
Another Feynman diagram for $Q^*(q) \to Q \overline
Q (P,\underline 8) + Q(p')$.

\item \label{fig3}
Feynman diagram for $Q^*(q) \to Q\bar q (P, \underline 1) + q (p')$.

\end{enumerate}

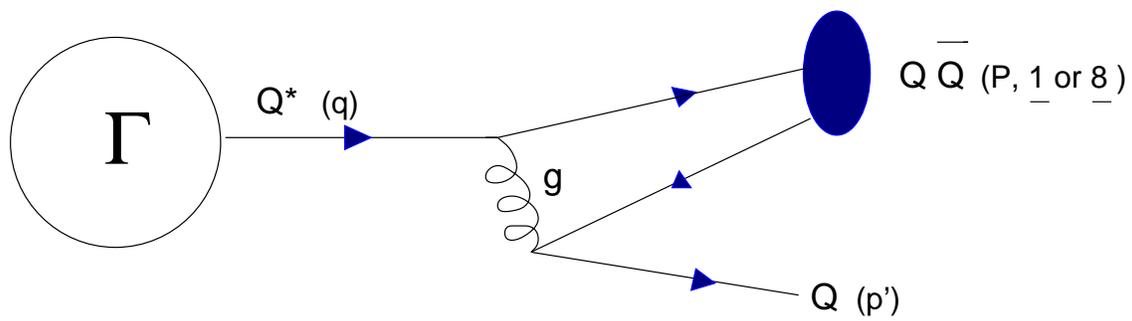
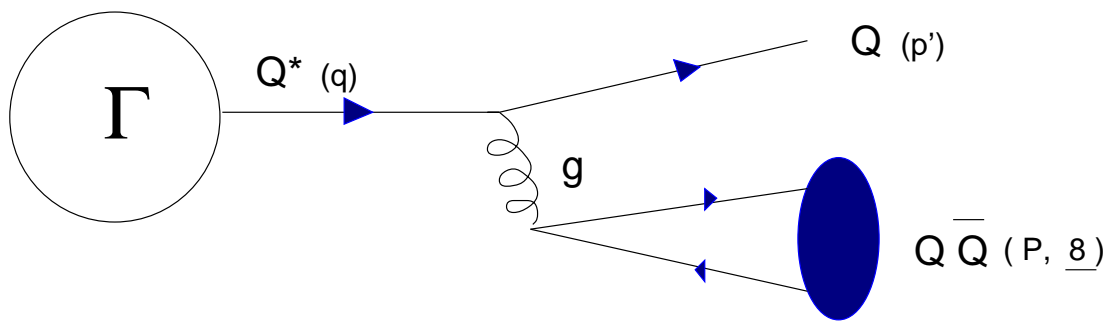
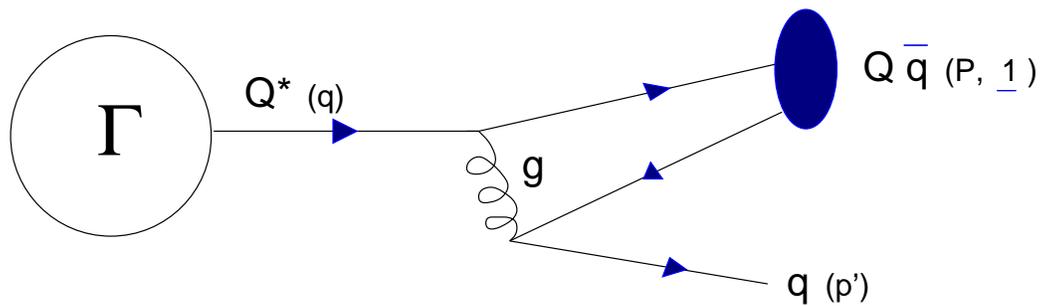
\end{document}